\documentclass[10pt,conference]{IEEEtran} 
\IEEEoverridecommandlockouts

\usepackage[table]{xcolor}
\usepackage{multirow,makecell}
\usepackage[most]{tcolorbox}
\usepackage{listings,amsfonts}
\usepackage{caption}
\usepackage{subcaption}
\usepackage{threeparttable}
\usepackage{bbding}
\usepackage{graphicx}
\usepackage{booktabs} 
\usepackage{longtable}
\usepackage{enumitem}

\definecolor{heatmap0}{HTML}{fafdcf}  
\definecolor{heatmap33}{HTML}{e8f6b1}
\definecolor{heatmap66}{HTML}{cfecb3}
\definecolor{heatmap90}{HTML}{c1e7b4}  
\definecolor{heatmap100}{HTML}{9cd8b8} 

\definecolor{customcite}{HTML}{b83b5e}
\definecolor{customlink}{HTML}{07689f}
\definecolor{customurl}{HTML}{1fab89}
\AtEndPreamble{
\usepackage{hyperref}

    \hypersetup{
      colorlinks = true,
      linkcolor = customlink,
      anchorcolor = purple,
      citecolor = customcite,
      filecolor = purple,
      urlcolor = customurl
    }
}

\definecolor{customcell}{HTML}{E6F2FF}

\def\BibTeX{{\rm B\kern-.05em{\sc i\kern-.025em b}\kern-.08em
    T\kern-.1667em\lower.7ex\hbox{E}\kern-.125emX}}

\newtcolorbox{takeawaybox}{
  colback=white,
  colframe=black,
  boxrule=0.6pt,
  arc=2.5pt,
  left=2pt,
  right=2pt,
  top=2pt,
  bottom=2pt,
  before skip=4pt,
  after skip=4pt
}

\begin{document}

\title{Unveiling the Landscape of LLM Deployment in the Wild: An Empirical Study}

\author{
\IEEEauthorblockN{Xinyi Hou\IEEEauthorrefmark{1}, Jiahao Han\IEEEauthorrefmark{1}, Yanjie Zhao, Haoyu Wang\IEEEauthorrefmark{2}}
\IEEEauthorblockA{
Huazhong University of Science and Technology, Wuhan, China\\
xinyihou@hust.edu.cn, jiahaohan789@gmail.com, yanjie\_zhao@hust.edu.cn, haoyuwang@hust.edu.cn}
\thanks{\IEEEauthorrefmark{1}Xinyi Hou and Jiahao Han contributed equally to this work.}
\thanks{\IEEEauthorrefmark{2}Haoyu Wang is the corresponding author (haoyuwang@hust.edu.cn).}
}

\maketitle

\begin{abstract}

Large language models (LLMs) are increasingly deployed through open-source and commercial frameworks, enabling individuals and organizations to self-host advanced LLM capabilities. As LLM deployments become prevalent, particularly in industry, ensuring their secure and reliable operation has become a critical issue. However, insecure defaults and misconfigurations often expose LLM services to the public internet, posing serious security and system engineering risks. 
This study conducted a large-scale empirical investigation of public-facing LLM deployments, focusing on the prevalence of services, exposure characteristics, systemic vulnerabilities, and associated risks. Through internet-wide measurements, we identified 320,102 public-facing LLM services across 15 frameworks and extracted 158 unique API endpoints, categorized into 12 functional groups based on functionality and security risk. Our analysis found that over 40\% of endpoints used plain HTTP, and over 210,000 endpoints lacked valid TLS metadata. API exposure was highly inconsistent: some frameworks, such as Ollama, responded to over 35\% of unauthenticated API requests, with about 15\% leaking model or system information, while other frameworks implemented stricter controls.
We observed widespread use of insecure protocols, poor TLS configurations, and unauthenticated access to critical operations. These security risks, such as model leakage, system compromise, and unauthorized access, are pervasive and highlight the need for a secure-by-default framework and stronger deployment practices.

\end{abstract}


\section{Introduction}

Driven by renowned models like OpenAI's GPT series~\cite{openai_gpt4} and DeepSeek's open-source variant~\cite{deepseek_llm}, large language models (LLMs) are rapidly gaining popularity and profoundly reshaping a wide range of applications. \textbf{Once primarily confined to research labs and industrial environments, these models are now not only continuously deployed in-depth within the industry, but are also gradually opened to the wider public, promoting the vigorous development of self-hosted and open source deployment.} The emergence of user-friendly tools and a vibrant community ecosystem~\cite{ollama, open_webui, anythingllm} has enabled individual enthusiasts, small enterprises, and developers to independently deploy and customize powerful LLMs for a variety of personal and professional needs, such as creative writing and content creation~\cite{isachenko2024generative}, software development and maintenance~\cite{hou2024large}, financial analysis and automated investment assistance~\cite{xie2024openfinllms}, and personal productivity tools, significantly enriching their daily digital experiences. However, as the barrier to entry for LLM deployment continues to decrease, an increasing number of deployments lack rigorous security considerations, exposing users and organizations to new operational and security risks.

Among these concerns, the OWASP Top 10 for LLM Applications 2025~\cite{owasp2025llm} identifies several risks that deserve particular attention during deployment, including sensitive information disclosure, unbounded consumption, and supply chain risks. These risks are especially relevant in self-hosted and open-source scenarios, where models, APIs, and supporting infrastructure are often exposed to the public internet without sufficient protection. These deployment challenges highlight not only the security concerns of LLM applications but also broader software engineering issues around system configuration, exposure management, and operational robustness.

Open-source frameworks commonly used for self-hosted LLM deployments often suffer from insecure default settings and misconfigurations, \textbf{exposing sensitive interfaces to the public Internet} without adequate protection and significantly enlarging the attack surface. Such services can be easily discovered via common asset-discovery tools like FOFA~\cite{fofa}, Shodan~\cite{shodan}, and ZoomEye~\cite{zoomeye}. For instance, Ollama~\cite{ollama}, a framework widely utilized for deploying local LLM services, exposes RESTful APIs publicly by default without authentication, enabling unauthorized operations such as model deletion, theft, GPU resource hijacking, and critical remote code execution (e.g., CVE-2024-37032~\cite{CVE-2024-37032}). Furthermore, publicly maintained platforms aggregating openly accessible Ollama services~\cite{freeollama} exacerbate the issue. Similarly, OpenWebUI~\cite{open_webui}, commonly integrated alongside Ollama for enhanced interaction capabilities, has suffered from vulnerabilities allowing arbitrary file uploads (CVE-2024-6707~\cite{CVE-2024-6707}), potentially facilitating remote command execution. Additionally, ComfyUI~\cite{comfyui}, known for its plugin-rich environment supporting diffusion-based generation tasks, has experienced multiple severe plugin-related security issues, including unauthorized remote code execution, arbitrary file access, and serialization vulnerabilities. Together, these vulnerabilities highlight the urgent need for stronger security awareness and practices in open-source LLM deployments.

\begin{figure*}[t!]
    \centering
    \includegraphics[width=0.99\linewidth]{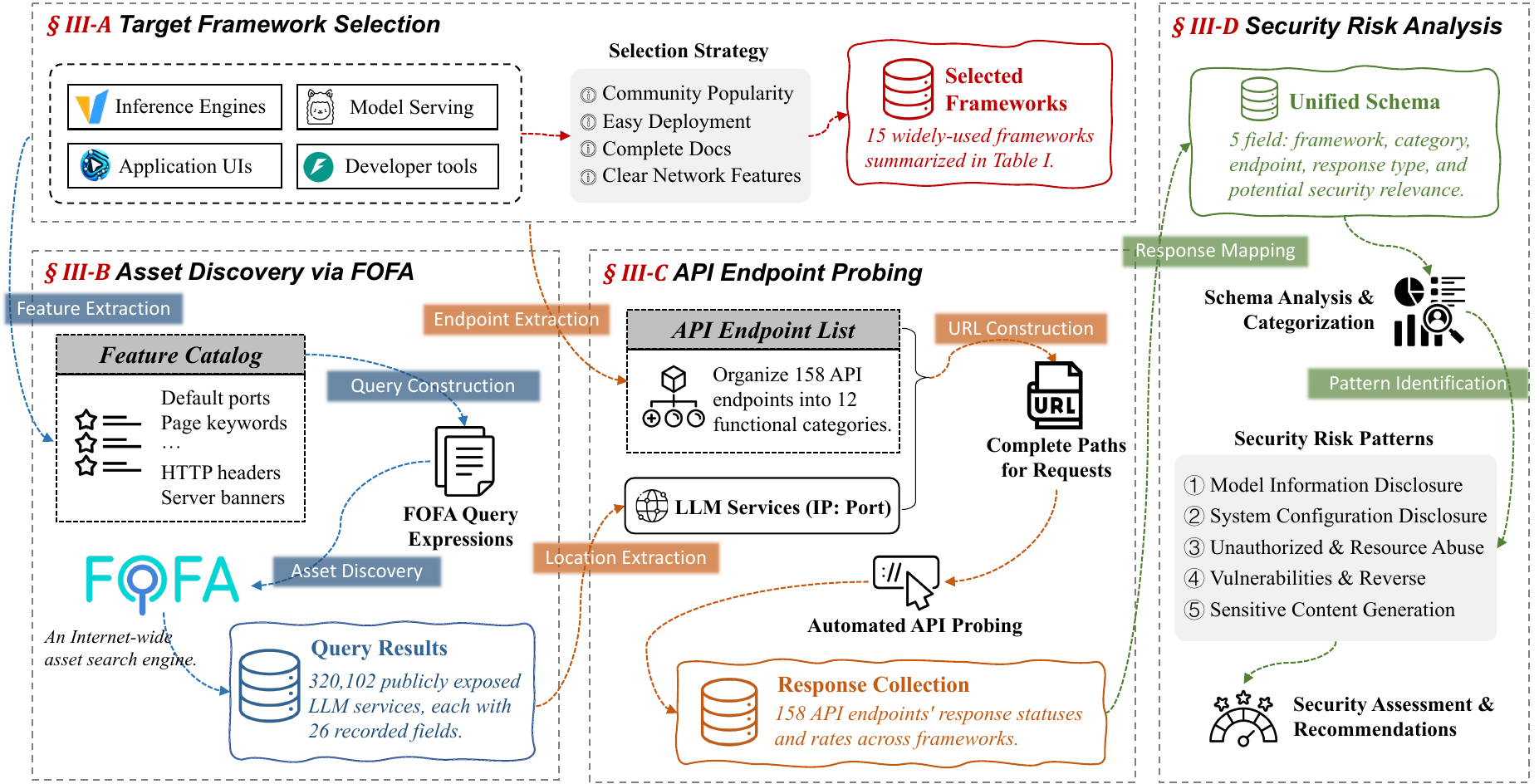}
    \caption{Overview of the Empirical Analysis Pipeline.}
    \label{fig:overview}
\end{figure*}

Motivated by these observations, we conduct a large-scale empirical study to examine the prevalence and characteristics of public-facing LLM deployments. 
Our investigation is guided by the following research questions (RQs):

\noindent\hangindent=2.5em\hangafter=1\textbf{RQ1 What is the global distribution and typical setup of public LLM services?}
We aim to map the current landscape of public LLM deployments, including their geographic, organizational, and technical characteristics.

\noindent\hangindent=2.5em\hangafter=1\textbf{RQ2 How do major LLM deployment frameworks differ in API exposure and access control?}
This helps us understand the diversity of deployment strategies and the effectiveness of default security mechanisms.

\noindent\hangindent=2.5em\hangafter=1\textbf{RQ3 What are the main security risks and vulnerabilities in public LLM deployments?}
We seek to identify prevalent threats and systemic issues that may affect the security of public LLM services.

\noindent To answer these RQs, our analysis pipeline is shown in \autoref{fig:overview}. We identify and analyze 320,102 public LLM services, spanning 15 popular deployment frameworks. From these, we extract 158 unique API endpoints, which we group into 12 functional categories based on their capabilities and associated security risks. Our study analyzes the geographic and network distribution of LLM deployments, uncovers widespread insecure configurations such as unauthenticated inference and model enumeration, and provides practical recommendations for developers and operators. The findings expose widespread security and deployment flaws, highlighting the need for secure-by-default designs and improved operational practices. 

\noindent\textbf{Contributions}. Our primary contributions\footnote{Our artifacts are publicly available at \url{https://github.com/cccccindy19/LLM-Services}.} are as follows:

\begin{itemize}
    \item \textbf{Large-Scale Empirical Study:} We identify 320,102 publicly accessible LLM services, providing the first empirical evidence of their global presence, deployment patterns, and exposure characteristics.
    
    \item \textbf{Exposure and Security Risk Analysis:} We systematically analyze 158 unique API endpoints, categorize them into 12 functional groups, and uncover systemic vulnerabilities in open-source LLM deployments.
    
    \item \textbf{Practical Recommendations:} We provide actionable guidance for secure-by-default design and deployment, grounded in empirical findings and targeting developers, framework maintainers, and operators.
\end{itemize}


\section{Background and Related Work} 
\label{sec: background}
\subsection{LLM Deployment Paradigms and Tooling}
An LLM framework provides the necessary foundation for building, deploying, and operating LLM-powered services in real-world environments. At a high level, an LLM deployment framework is a software platform or toolkit that enables organizations to efficiently and securely operate LLMs and connect models with users and applications at scale. Such frameworks typically integrate multiple layers of computing, service management, user interaction, and developer support.

This stack broadly consists of four key components.
\textbf{Inference Engines.}  
Inference engines execute LLM computations across diverse hardware. Optimized systems like vLLM~\cite{vllm} and Hugging Face Transformers leverage techniques such as continuous batching, KV cache optimization, and operator fusion to improve throughput and latency~\cite{li2024llm}. Tools like LLM-Pilot~\cite{lazuka2024llm} support benchmarking and tuning in production environments.
\textbf{Model Serving Frameworks.}  
Serving frameworks manage hosting, request routing, and scaling. Systems like Ray Serve~\cite{ray_serve} and Ollama~\cite{ollama} support flexible deployment across cloud and edge platforms~\cite{chen2025agile, bambhaniya2024demystifying}. Split-based deployment further distributes model components to balance performance and privacy~\cite{chen2024unveiling}.
\textbf{Application User Interfaces.}  
User-facing applications integrate LLMs via conversational interfaces, RAG pipelines, and autonomous agents. Frameworks like LangChain and ChatGPT plugins simplify this integration but also introduce risks such as indirect prompt injection~\cite{wu2024wipi, huq2023s}.
\textbf{Developer Tools and Ecosystems.}  
Developer tools support fine-tuning, evaluation, and maintenance. Libraries such as DeepSpeed, Hugging Face PEFT, and LoRA enable efficient model adaptation~\cite{hou2025next, shetty2024building}, while tools like TestGen-LLM~\cite{alshahwan2024automated} automate engineering workflows. Specialized accelerators are also emerging to meet large-scale inference demands~\cite{firouzi2025aid, chen2025agile}.
Despite growing modularity and accessibility, little is known about how LLM deployments are configured and exposed in practice. To bridge this gap, we conduct the first large-scale analysis of publicly accessible LLM services in the wild.

\subsection{Security Challenges in LLM Services}

Large-scale LLM deployments introduce security challenges across networking, authentication, input handling, and resource management layers. A common risk stems from accidental exposure, where open ports or permissive configurations allow unrestricted access to LLM services. Measurements of Internet-facing deployments~\cite{wu2024wipi} reveal widespread misconfigurations, leading to unauthorized interactions, resource abuse, and data leakage.
Authentication weaknesses further exacerbate these risks. Pesati et al.~\cite{pesati2024security} found that many public LLM services lack proper authentication or rely on fragile mechanisms, exposing them to prompt manipulation, session hijacking, and privilege escalation.
Input manipulation, particularly prompt injection, presents another critical threat. Indirect injections, where malicious inputs are embedded in external content, can subvert model behavior without user awareness~\cite{wu2024wipi, greshake2023not, huq2023s}. Attackers can exploit LLM integrations with web tools, APIs, or retrieval systems to trigger unauthorized actions or leak data~\cite{mohammad2024web, chiang2025web}.
Multi-tenant serving architectures introduce side-channel vulnerabilities. Sharing Key-Value (KV) caches among users, as in vLLM, can leak cross-tenant information. Attacks like PROMPTPEEK~\cite{wu2025know} demonstrate that adversaries can reconstruct other users' prompts by analyzing cache access patterns.
Furthermore, integrating LLMs with plugins, autonomous agents, and retrieval-augmented generation (RAG) pipelines significantly expands the attack surface~\cite{wu2024new, yao2025controlnet, li2025commercial}. Vulnerabilities in these systems can lead to data poisoning, unauthorized API calls, and model evasion. Surveys~\cite{yao2024survey, he2024emerged} highlight that LLM-based agents are especially prone to security and privacy threats due to their complex interactions with external systems.
Motivated by these risks, we systematically investigate real-world LLM deployments to uncover prevalent insecure practices and highlight critical gaps in existing security postures.

\section{Methodology}
\label{sec:methodology}

\autoref{fig:overview} outlines our four-step methodology for analyzing LLM deployments in the wild. We first select representative deployment frameworks (\autoref{subsec:frameworks}), then discover publicly accessible instances via FOFA (\autoref{subsec:fofa}), probe their APIs to collect metadata (\autoref{subsec:endpoint}), and finally analyze their configurations and security posture (\autoref{subsec:analysis}).

\subsection{Target Framework Selection}
\label{subsec:frameworks}

To structure our measurement and ensure broad coverage across the LLM deployment stack, we selected representative frameworks across four functional categories: \textbf{Inference engines} (e.g., vLLM~\cite{vllm}, llama.cpp~\cite{llama_cpp}) handle local model execution optimized for performance. \textbf{Model serving frameworks} (e.g., Ollama~\cite{ollama}, Ray Serve~\cite{ray_serve}) expose scalable APIs. \textbf{Application UIs} (e.g., Open WebUI~\cite{open_webui}, Jan~\cite{jan}, ComfyUI~\cite{comfyui}) provide user-facing interfaces, while \textbf{developer tools} (e.g., Jupyter Notebook~\cite{jupyter}, FastAPI~\cite{fastapi}) facilitate development workflows and integration.
Our framework selection followed a multi-step process: First, we surveyed the most cited projects in open source communities (e.g., GitHub, Hugging Face) and industry forums to compile an initial shortlist. We then screened these frameworks based on factors such as ease of deployment (e.g., availability of Docker images or installation scripts), quality and completeness of documentation, and evidence of active maintenance (recent commits, issue activity). Next, we prioritized frameworks that exhibited unique network exposure patterns and supported diverse deployment scenarios, such as on-premises, cloud, and hybrid deployments. This process enabled us to capture both widely adopted solutions and emerging tools with unique characteristics. In total, we selected 15 widely used frameworks, which are summarized in \autoref{tab:llm_deployment_frameworks}.

\begin{table*}[htbp]
\centering
\caption{Publicly Exposed LLM Deployment Frameworks Discovered via FOFA (As of April 20, 2025).}
\label{tab:llm_deployment_frameworks}
\resizebox{1\linewidth}{!}{
\begin{tabular}{  
    c   
    l   
    l   
    l   
    r   
}
\toprule[1.2pt]
\textbf{Category}                & \textbf{Framework}           & \textbf{Description}                                       & \textbf{Key Feature}                    & \textbf{Count} \\
\midrule[1.2pt]

\multirow{4}{*}{Inference Engines}       
                                 & vLLM~\cite{vllm}         & High‑throughput, memory‑optimized inference service         & title/vLLM; port=8000                   & 6,077   \\
                                 & llama.cpp~\cite{llama_cpp}    & C/C++ quantized inference engine                            & title/llama.cpp; port=8080              & 4,234   \\
                                 & GPT4All~\cite{gpt4all}      & Local GPT model runtime exposing REST endpoint              & title/GPT4All; port=8080                & 2,572   \\
                                 & Llamafile~\cite{llamafile}    & Single‑file GPT execution tool                              & title/Llamafile; port=8080              & 39      \\

\midrule
\multirow{3}{*}{Model Serving}  
                                 & Ollama~\cite{ollama}       & Cross‑platform CLI for local LLM API service                & header/Ollama; port=11434               & 155,423 \\
                                 & AnythingLLM~\cite{anythingllm}  & Local knowledge base integration with LLM API               & title/AnythingLLM; port=3000            & 3,766   \\
                                 & Ray Serve~\cite{ray_serve}    & Scalable microservice framework with autoscaling            & title/Ray Serve; port=8000              & 365     \\

\midrule
\multirow{6}{*}{Application UIs}    
                                 & Open WebUI~\cite{open_webui}   & Ollama/GPT‑API Web dashboard                                & title/Open WebUI; port=8080             & 37,242  \\
                                 & Jan~\cite{jan}          & Interactive local chat UI                                   & title/Jan; port=3000                    & 28,445  \\
                                 & NextChat~\cite{nextchat}     & Local ChatGPT‑style interface                               & title/chatgpt‑next‑web; port=3000       & 25,883  \\
                                 & ComfyUI~\cite{comfyui}      & Visual workflow builder for LLM and image pipelines         & title/ComfyUI; port=8188                & 15,219  \\
                                 & Gradio~\cite{gradio}       & Python toolkit for shareable LLM web demos                  & title/Gradio; port=7860                 & 9,729   \\
                                 & Text Generation Web UI~\cite{textgen_webui} & Generic LLM Web interface                          & title/text‑generation‑webui; port=7860  & 2,051   \\

\midrule
\multirow{2}{*}{Developer Tools} 
                                 & Jupyter Notebook~\cite{jupyter}    & Interactive Python notebook environment                     & body/Jupyter Notebook                   & 24,531  \\
                                 & FastAPI/Swagger UI~\cite{fastapi}  & Auto‑generated API docs and interactive endpoints           & title/FastAPI; port=8000                & 4,526   \\
\midrule
\multicolumn{4}{l}{\textbf{Total}} & \textbf{320,102} \\
\bottomrule[1.2pt]
\end{tabular}}
\end{table*}

\subsection{Asset Discovery via FOFA}
\label{subsec:fofa}

We use FOFA, a widely adopted Internet-wide asset search engine that indexes IP addresses, domains, and service metadata~\cite{fofa}, to discover public-facing LLM deployments at scale.
For each tool, we built a feature catalog that captures its unique network-level characteristics, such as default port (e.g., Ollama's default port is 11434), page title or HTML keywords (e.g., \texttt{title=Open WebUI''}), HTTP headers (e.g., \texttt{Ollama is running''}), favicon hashes, and known API paths (see \autoref{tab:llm_deployment_frameworks} for details). 
Based on this catalog, we designed custom FOFA query expressions using the platform’s advanced syntax. For instance, \texttt{app=Ollama'' \&\& port=11434''} for Ollama and \texttt{title=FastAPI'' || body=FastAPI'' \&\& (port=8000'' || port=8080'')} for FastAPI).
To improve coverage and reduce false positives, we refined these queries iteratively. This process involved testing against known deployment samples, consulting official documentation, and manually validating a representative subset of results. In particular, we examined rendered landing pages, HTTP headers, and available API metadata to confirm the identity of suspected services.
As of April 20, 2025, we had identified 320,102 publicly accessible LLM-related services across 15 representative deployment tools, as shown in \autoref{tab:llm_deployment_frameworks}. These results form the foundation for the subsequent phases of our study, including endpoint probing and configuration analysis.

\subsection{API Endpoint Probing}
\label{subsec:endpoint}

We extracted 158 API endpoints from the official documentation of 15 widely used LLM deployment frameworks. These endpoints span many functionalities, including model management, file access, kernel and session management, and application deployment. As summarized in \autoref{tab:endpoint}, the endpoints are organized into 12 functional categories. For instance, endpoints under the \textit{Text/Chat} category enable OpenAI-style text generation (e.g., \texttt{/v1/chat/completions}), while the \textit{Model Control} group provides access to model lifecycle operations such as loading or deleting models (e.g., \texttt{/models/delete}). \textbf{Several endpoints were found to expose potentially sensitive actions, including file uploads, session control, and even server-side code execution in certain configurations.}
Using the extracted paths, we combined them with the network locations of previously identified LLM services to construct complete URLs. An automated process then probed these URLs with HTTP(S) requests, systematically collecting and storing all responses for further analysis. The resulting dataset provides a comprehensive view of exposed API functionalities and their response behaviors in real-world deployments.

\begin{table}[h!]
\centering
\caption{API Categories in Exposed LLM Services.}
\resizebox{1\linewidth}{!}{
\begin{tabular}{rlr}
\toprule[1.2pt]
\textbf{Category} & \textbf{Function} & \textbf{\#} \\ \midrule[1.2pt]
Text \& Chat Gen     & Text and chat completions (OpenAI-style)          & 23 \\
Embedding Gen        & Generate vector embeddings from text              & 9  \\
Image \& Audio       & Image generation, editing, speech-to-text, TTS    & 32 \\
Model Ops            & Load, list, delete, and inspect models            & 28 \\
File Ops             & Upload, download, or delete general files         & 19 \\
Knowledge/RAG        & Upload and query knowledge bases for RAG          & 3  \\
Fine-tuning Tasks    & Create fine-tuning jobs and upload training data  & 10  \\
Session \& Kernel    & Manage Jupyter sessions, kernels, and clusters    & 16 \\
Task Queue           & Submit tasks, monitor queues, and job scheduling  & 3  \\
System Config        & Get system status, API version, and configuration & 6  \\
Moderation Check     & Content safety and moderation checks              & 5  \\
App Deployment       & Deploy or manage LLM apps (e.g., Ray Serve)       & 4  \\ 
\midrule
\textbf{Total}       &                                                     & 158 \\
\bottomrule[1.2pt]
\end{tabular}}
\label{tab:endpoint}
\end{table}

\subsection{Configuration and Security Analysis}
\label{subsec:analysis}

Following endpoint probing, we analyzed the returned API responses to assess the configuration and security posture of exposed LLM services. Each response was mapped to a unified schema capturing five key fields: \textit{deployment framework}, \textit{endpoint category}, \textit{endpoint path}, \textit{response type} (e.g., success, denial, error), and \textit{potential security relevance}. This schema enabled consistent comparison across frameworks with heterogeneous designs and naming conventions.
The analysis focused on endpoint responsiveness under unauthenticated access, functional exposure across 12 categories, and the presence of risky operations such as model management, file access, and code execution. 

\section{Results}
\label{sec:results}

This section presents the results of our empirical analysis of public LLM services. Findings are organized by RQs, covering overall exposure, deployment characteristics, configuration patterns, and security risks observed in the dataset.

\subsection{RQ1: General Statistics}

We begin with basic statistical analyses to summarize the global deployment, organizational involvement, exposure surfaces, and security postures of self-hosted LLM services.

\subsubsection{Global and Organizational Deployment Trends}

\begin{figure}[htbp]
    \centering
    \begin{subfigure}[b]{0.49\linewidth}
        \centering
        \includegraphics[width=\linewidth]{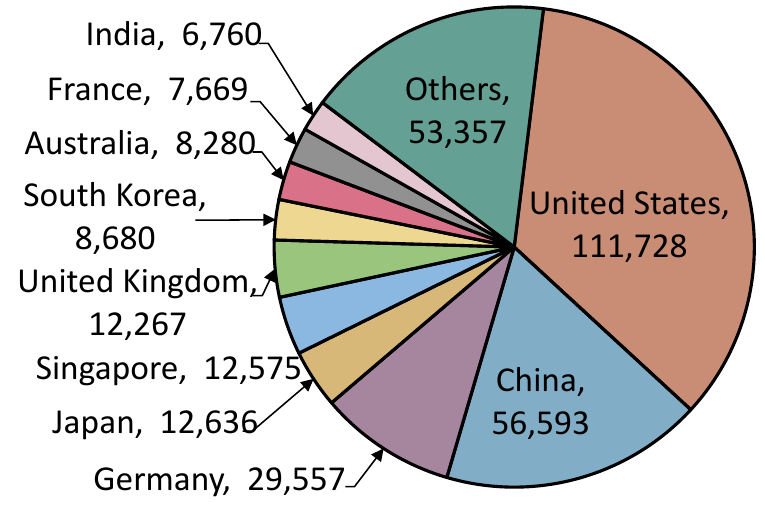}
        \caption{Geographic Distribution.}
        \label{fig:country}
    \end{subfigure}
    \hfill
    \begin{subfigure}[b]{0.49\linewidth}
        \centering
        \includegraphics[width=\linewidth]{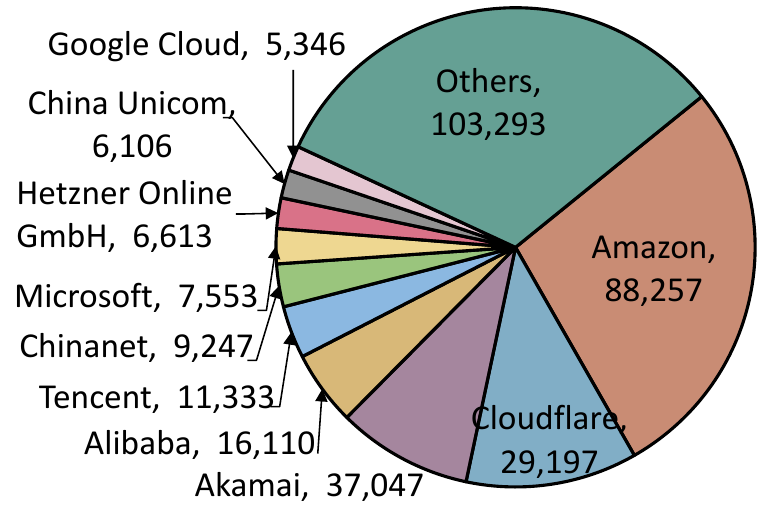}
        \caption{Organization Distribution.}
        \label{fig:org}
    \end{subfigure}
    \caption{Global Landscape of Public LLM Deployment.}
    \label{fig:global}
\end{figure}

We analyze 320,012 exposed LLM service endpoints, aggregating their origin by country and associated hosting organization. \autoref{fig:global} illustrates this landscape with two pie charts summarizing the top contributors by country and organization.

\noindent\textbf{Geographic Centralization.}
As shown in \autoref{fig:country} , the United States leads with 111,728 public LLM services, more than double the number of second-place China (56,593). Other prominent contributors include Germany, Japan, and Singapore. This imbalance highlights a pronounced centralization of deployment activity within technologically advanced nations, likely driven by both cloud infrastructure maturity and early AI adoption.
Notably, the ``Others'' category still accounts for over 53,000 instances, suggesting that self-hosted LLMs are also gaining traction in the broader global community, albeit in smaller clusters. This distribution hints at a growing democratization of LLM capabilities, though one is still heavily shaped by infrastructure access and national regulatory frameworks.

\noindent\textbf{Organizational Dominance and Long Tail.}
As shown in \autoref{fig:org}, a few major providers dominate public LLM deployments. Amazon alone hosts 88,257 instances, followed by Cloudflare, Akamai, and Microsoft. This concentration is not coincidental: these cloud and CDN providers offer highly accessible and scalable infrastructure, often with free-tier or pay-as-you-go models, making them attractive to both individual developers and small organizations. In many cases, LLM frameworks are also preconfigured for platforms like AWS or Cloudflare Workers, further lowering deployment barriers.
Interestingly, a long-tail distribution persists: over 100,000 services are hosted by entities categorized as ``Others''. This indicates that a substantial number of deployments originate from smaller cloud vendors, university networks, hobbyist servers, or edge nodes. While this decentralization reflects the democratization of LLM deployment, it also introduces significant heterogeneity in operational practices.

This uneven deployment landscape creates a strategic asymmetry: while \textbf{centralized platforms enable efficient patching and policy enforcement, the fragmented long tail poses significant challenges due to inconsistent security practices}. Attackers can exploit this imbalance by focusing on a few high-density targets, whereas defenders must cope with a much broader and less predictable surface.

\subsubsection{High-Traffic Domains and Service Concentration}

More than 210,000 exposed LLM services lack valid domain assignments, representing a significant portion of the overall deployment landscape. This reflects widespread deployment without proper DNS configuration or certificate binding, likely resulting from incomplete setup processes or reliance on default settings in automated toolchains. These services are typically accessible only via IP address and represent a considerable portion of deployments with poor post-deployment hygiene, weakening traceability and trust mechanisms.

\begin{figure}[htbp]
    \centering
    \includegraphics[width=0.95\linewidth]{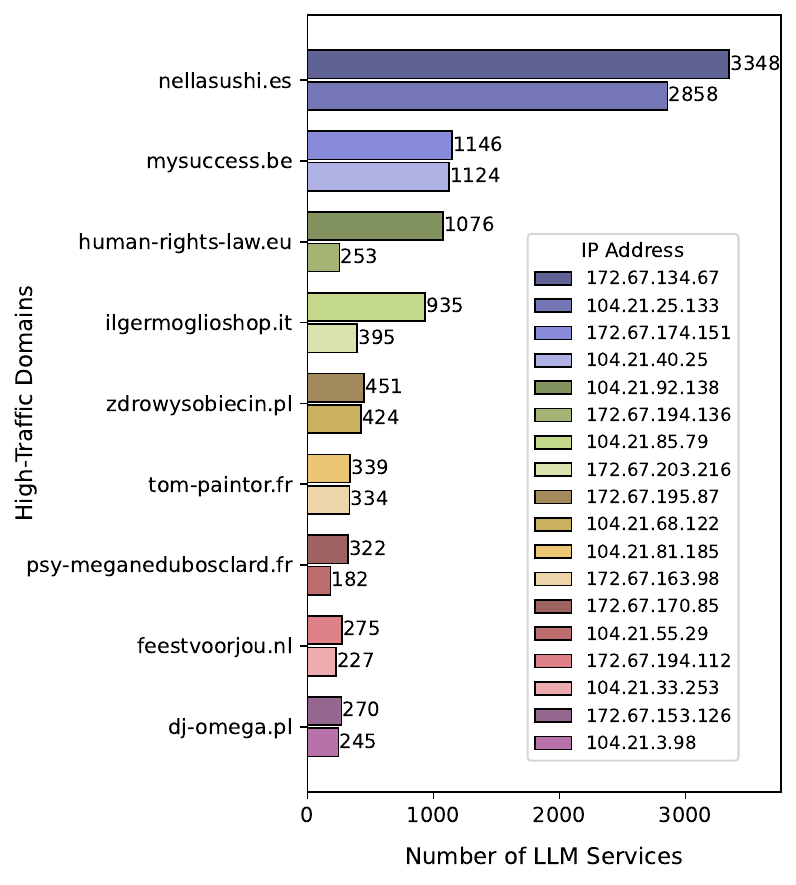}
    \caption{High-Traffic Domains by Number of LLM Services.}
    \label{fig:domain}
\end{figure}

Beyond these unassigned cases, certain domains are associated with an unusually high number of LLM instances. As shown in \autoref{fig:domain}, domains such as \texttt{nellasushi.es}, \texttt{mysuccess.be}, and \texttt{human-rights-law.eu} each host thousands of services. This level of repetition is uncommon in conventional web deployments and suggests that these domains serve as default endpoints in automated or templated hosting environments. Two factors support this interpretation. First, services under the same domain frequently share a small number of IP addresses, indicating centralized hosting or large-scale reuse of identical deployment images. For instance, 6,206 instances under \texttt{nellasushi.es} are served by two IPs. Second, deployment metadata\footnote{Supplemental heatmap available at \url{https://github.com/cccccindy19/LLM-Services}.} reveals a consistent reliance on a limited set of frameworks such as ComfyUI, Jan, and vLLM, often in their default configurations. These patterns point to widespread use of prebuilt containers or orchestration scripts.
Such concentration has both operational advantages and security implications. Focusing remediation efforts on a few high-frequency domains could reduce exposure at scale. However, \textbf{any misconfiguration or compromise within these clusters could simultaneously affect thousands of endpoints}. Moreover, the repeated use of domain names, IPs, and certificates erodes the reliability of trust models and complicates service attribution.

\begin{figure*}[t!]
    \centering
    \includegraphics[width=1\linewidth]{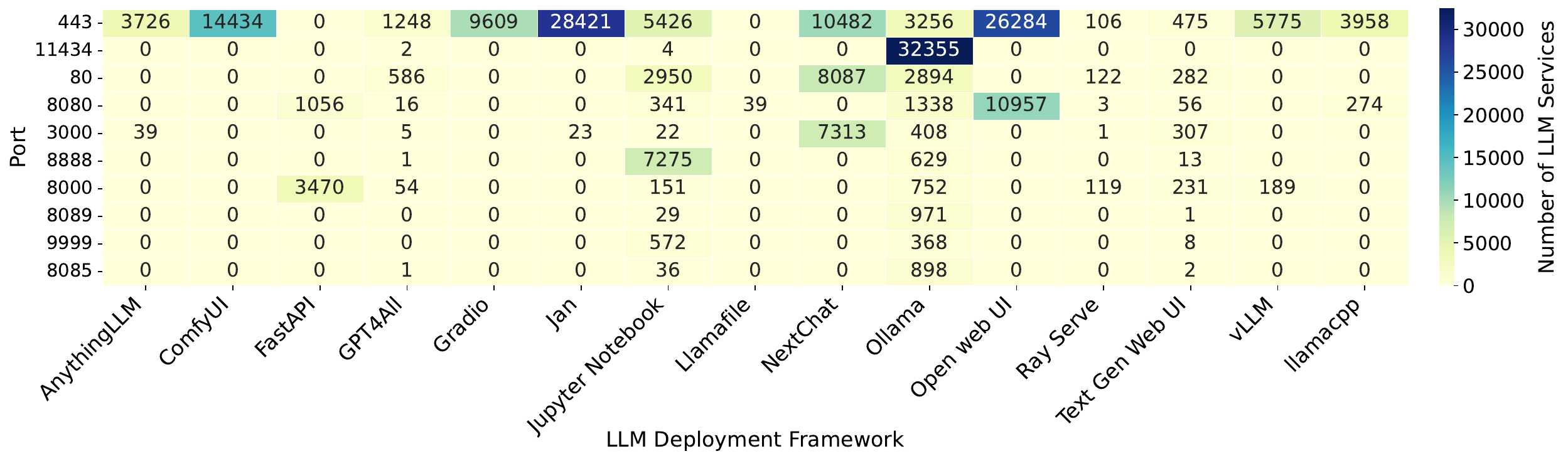}
    \caption{Distribution of LLM Services Across Ports by Deployment Framework.}
    \label{fig:port_heatmap}
\end{figure*}

\subsubsection{Server Stack Composition}

Public LLM services are predominantly hosted using familiar and widely adopted server stacks. As shown in \autoref{tab:os_server}, the most common configuration is \texttt{Ubuntu + nginx}, with versions such as \texttt{nginx/1.18.0} and \texttt{1.24.0} being particularly prevalent. Apache-based deployments are also observed, primarily on Debian or generic Unix systems, while Microsoft IIS accounts for only a small fraction, reflecting a strong bias toward Linux-based environments.
However, a substantial number of services lack complete environment metadata. Over 50,000 instances do not report a valid operating system, and more than 90,000 specify only generic server identifiers. These incomplete configurations likely arise from containerized or scripted deployments where base image information is obscured. While such approaches streamline deployment, they also reduce observability and hinder vulnerability assessment.
Some configurations also raise potential security concerns. Instances running outdated server versions (e.g., \texttt{nginx/1.14.0}, \texttt{apache/2.4.29}) may expose known vulnerabilities if not properly patched~\cite{nginx_cve,apache_cve}. Lightweight servers such as \texttt{Uvicorn} and \texttt{TornadoServer}, though efficient, often lack default hardening and are more prone to insecure defaults. The prevalence of opaque and minimalist setups further complicates automated auditing and increases uncertainty in assessing the broader exposure surface.

\begin{table}[h!]
\centering
\caption{Top OS–Server Deployment Combinations in Exposed LLM Services.}
\resizebox{1\linewidth}{!}{
\begin{tabular}{clcrr}
\toprule[1.2pt]
\textbf{OS} & \textbf{Server} & \textbf{Count} & \textbf{Percentage} & \textbf{Share in OS} \\ \midrule
ubuntu  & nginx/1.18.0 (ubuntu)        & 7,242 & 42.84\% & 53.23\% \\
ubuntu  & nginx/1.24.0 (ubuntu)        & 3,659 & 21.65\% & 26.9\%  \\
windows & microsoft-iis/10.0           & 567  & 3.35\%  & 50.04\% \\
ubuntu  & apache/2.4.52 (ubuntu)       & 545  & 3.22\%  & 4.01\%  \\
debian  & apache/2.4.62 (debian)       & 457  & 2.70\%  & 44.63\% \\
ubuntu  & apache/2.4.41 (ubuntu)       & 420  & 2.48\%  & 3.09\%  \\
ubuntu  & nginx/1.14.0 (ubuntu)        & 389  & 2.30\%  & 2.86\%  \\
ubuntu  & apache/2.4.29 (ubuntu)       & 357  & 2.11\%  & 2.62\%  \\
unix    & apache/2.4.62 (unix)         & 218  & 1.29\%  & 26.42\% \\
ubuntu  & apache/2.4.58 (ubuntu)       & 194  & 1.15\%  & 1.43\%  \\
ubuntu  & nginx/1.26.0 (ubuntu)        & 194  & 1.15\%  & 1.43\%  \\
debian  & apache/2.4.25 (debian)       & 139  & 0.82\%  & 13.57\% \\
unix    & apache/2.4.57 (unix)         & 122  & 0.72\%  & 14.79\% \\
windows & microsoft-iis/8.5            & 119  & 0.70\%  & 10.5\%  \\
unix    & apache/2.4.63 (unix)         & 117  & 0.69\%  & 14.18\% \\
ubuntu  & nginx/1.10.3 (ubuntu)        & 112  & 0.66\%  & 0.82\%  \\
windows & microsoft-iis/7.5            & 101  & 0.60\%  & 8.91\%  \\
\bottomrule[1.2pt]
\end{tabular}}
\label{tab:os_server}
\end{table}

\subsubsection{Communication Security and Port Exposure}

\begin{figure}[htbp]
    \centering
    \includegraphics[width=1\linewidth]{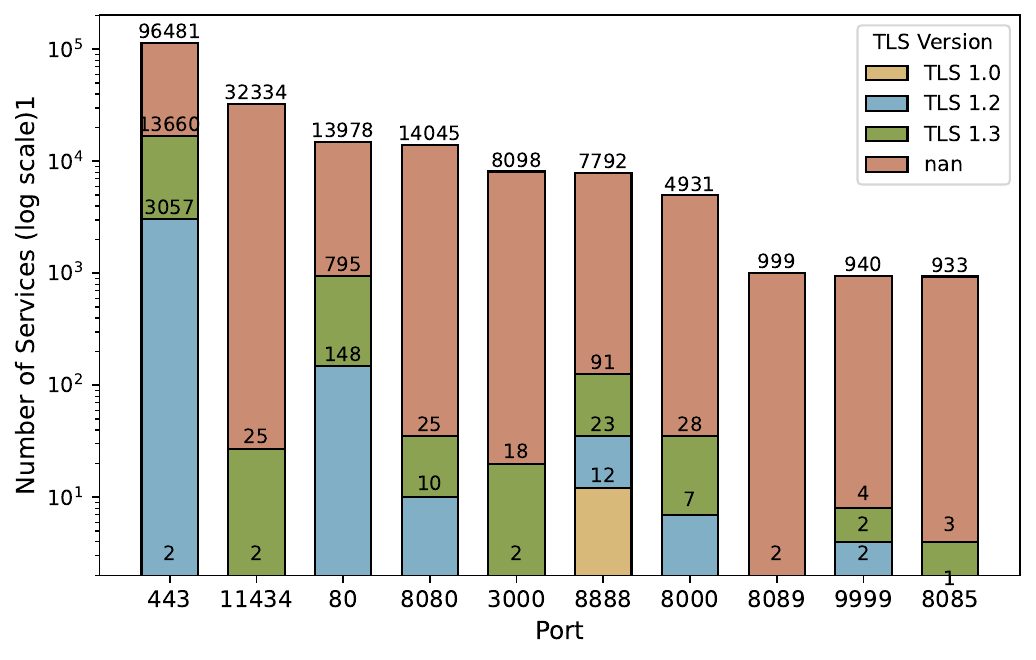}
    \caption{TLS Version Distribution Across Ports (Log Scale).}
    \label{fig:port}
\end{figure}

The communication security of public LLM services varies significantly across ports and deployment frameworks. As shown in \autoref{fig:port}, a large fraction of services either lack TLS entirely or use outdated versions. Port 443, typically associated with HTTPS, is the most secure, with over 13,000 instances using TLS 1.3 and only 2 instances using TLS 1.0. However, this is not the norm across other ports.
Overall, \textbf{nearly 129,811 services are still accessible via plain HTTP, accounting for over 40\% of the measured endpoints}. This reflects a broad lack of transport encryption across the LLM service ecosystem. Notably, a large number of services on ports such as 11434, 3000, and 8888 lack TLS support. These ports are commonly associated with frameworks including Open Web UI, Jan, and Ollama, as shown in \autoref{fig:port_heatmap}. For instance, over 32,000 services on port 11434 and nearly 14,000 on 8080 lack any TLS configuration. Even when TLS is present, a non-trivial portion still relies on deprecated versions such as TLS 1.0 and 1.2, which expose services to downgrade attacks and known cryptographic vulnerabilities.

The widespread use of non-standard ports, often with weak or missing encryption, highlights a broader issue: many LLM frameworks prioritize ease of deployment over secure defaults. This creates a communication surface that is both predictable and exposed. Attackers can scan for specific ports tied to popular frameworks and exploit weak encryption to intercept or tamper with LLM outputs.
These findings highlight the need for stronger default security in deployment tools, especially in TLS enforcement and port hardening. Without it, even well-intentioned self-hosted setups remain vulnerable by design.

\subsubsection{Certificate Reuse and Identity Management}

\begin{figure}[t!]
    \centering
    \includegraphics[width=1\linewidth]{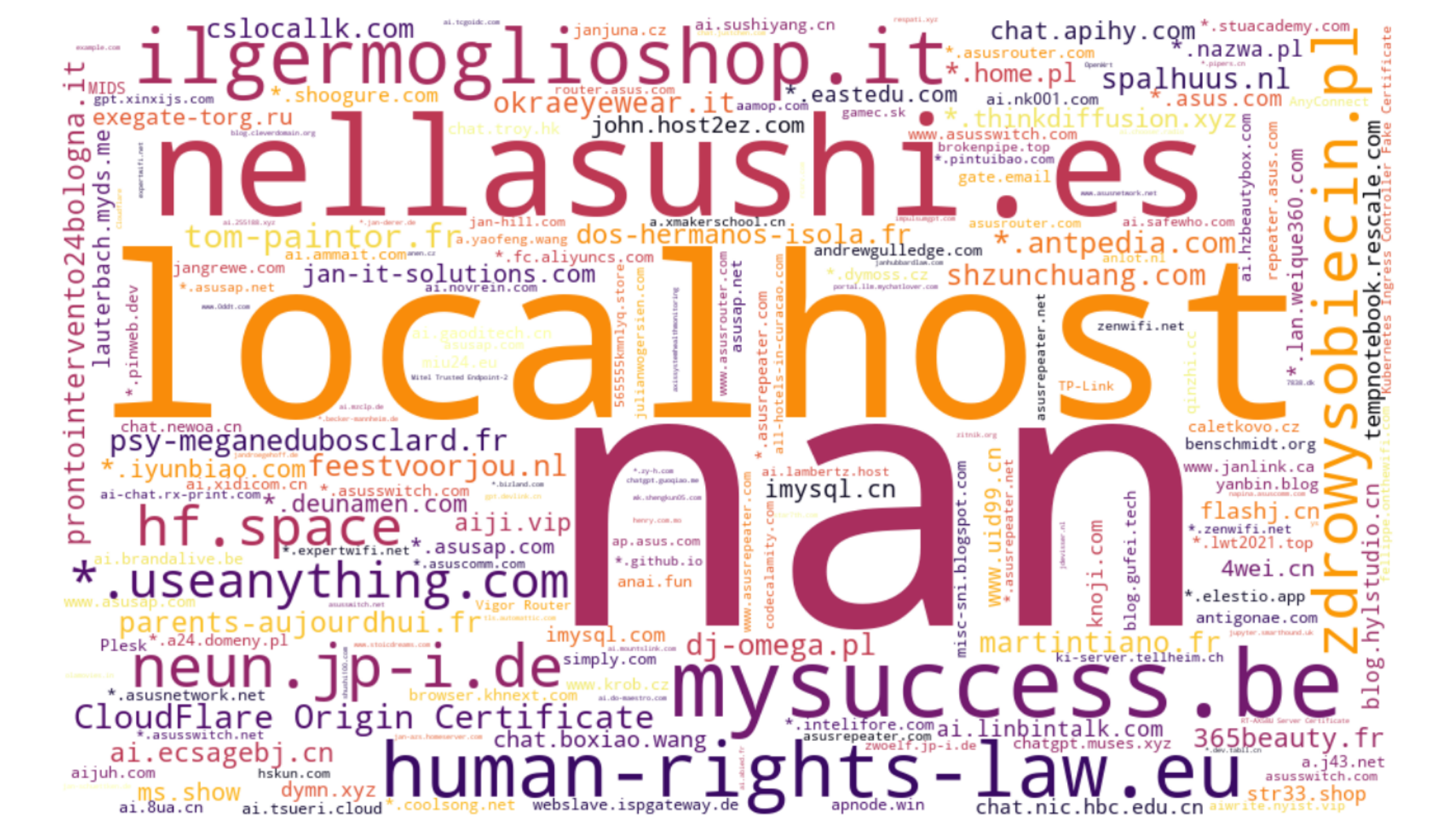}
    \caption{Distribution of Reused TLS Certificate Subject CNs.}
    \label{fig:cert}
\end{figure}

TLS certificate metadata associated with public LLM services reveals widespread reuse and misconfiguration, especially in certificate subject fields. As illustrated in \autoref{fig:cert}, common names (CNs) such as \texttt{localhost} and generic domain labels (e.g., \texttt{nellasushi.es}, \texttt{mysuccess.be}) appear frequently, while a significant number of entries are simply marked as \texttt{nan}, indicating missing or unparsable certificate subject fields.
Quantitative analysis supports this observation. Over 210,000 services use \texttt{localhost} or \texttt{nan} as the certificate subject CN, and many others share identical organization names across hundreds or thousands of instances. While some reuse is expected in containerized or replicated environments, the scale observed here suggests a lack of certificate management hygiene. This undermines the integrity of TLS-based identity validation, as the same certificate may appear in unrelated deployments or across unaffiliated IP ranges.

Issuer fields show strong centralization, with certificates mostly issued by \texttt{Cloudflare}, \texttt{Let's Encrypt}, and \texttt{ZeroSSL}, often via automated pipelines. However, they are often paired with missing or generic subject metadata, weakening endpoint authenticity.
TLS fingerprinting analysis via \texttt{ja3s} values reveals limited diversity, with a few signatures covering most services. This suggests many deployments use default TLS configurations, making them more vulnerable to fingerprint-based traffic correlation or blocking.
Together, these patterns expose a weak identity layer in the self-hosted LLM ecosystem, undermining encryption’s role in both confidentiality and endpoint authentication.

\begin{takeawaybox}
\textbf{Answer to RQ1:} \textit{Public LLM deployments are spread across the globe, with most exposure concentrated on a few major providers (e.g., Amazon, Cloudflare), but also encompassing many smaller networks. Of the 320,012 endpoints, over 40\% use plain HTTP, and over 210,000 lack valid TLS metadata. Deployment setup and security practices are highly inconsistent, highlighting the need for standardization and secure-by-default practices.
}
\end{takeawaybox}

\subsection{RQ2: API Exposure}

Building on the exposure landscape, we examine API responsiveness across frameworks, coverage of functional categories, and the structure of per-endpoint results.

\subsubsection{Framework-Level API Responsiveness}

We observed significant variation in API responsiveness among publicly available LLM deployments. As shown in \autoref{tab:response}, frameworks such as Ollama and Llamafile responded to over 10\% of unauthenticated API requests, indicating permissive defaults and minimal access controls. In contrast, platforms such as Open WebUI, NextChat, and Gradio had response rates below 0.1\%, suggesting more stringent backend protection or exposing only the frontend. FastAPI and AnythingLLM were excluded due to their lack of standardized and comparable APIs. 
Due to the large number of Ollama deployments, we conducted API probing on a small random sample of 877 instances. As shown in the first row of \autoref{tab:response} , 35.23\% of Ollama instances responded. However, upon closer inspection, we discovered that a significant portion of these responses consisted of cookie-cutter ``template responses'' (generic, uninformative replies) rather than meaningful API output. We speculate that these template responses may be due to default configurations, placeholder endpoints, or defensive mechanisms designed to prevent automated probing. Excluding template responses reduced the actual success rate to 15.28\%, closely matching the overall response rate (14.69\%) observed in the full Ollama dataset of 155,423 instances. Notably, we did not observe such template response patterns in other frameworks.

\begin{table}[htbp]
\centering
\caption{Responsiveness of Deployment Frameworks.}
\resizebox{1\linewidth}{!}{
\begin{threeparttable}
\begin{tabular}{cccr}
\toprule[1.2pt]
\textbf{Framework} & \textbf{\# Resp.} & \textbf{Sample / Total} & \textbf{Resp. Rate} \\
\midrule[1.2pt]
\multirow{2}{*}{Ollama} & 309 (w/)\tnote{1} & 877 & \cellcolor{heatmap100}35.23\% \\
 & 134 (w/o)\tnote{2} & 877 & \cellcolor{heatmap33}15.28\% \\\midrule
Ollama & 22{,}831 & 155{,}423 & \cellcolor{heatmap33}14.69\% \\ 
Open WebUI & 10 & 43{,}947 & 0.02\%  \\
Jan & 7 & 380 & 1.84\%  \\
NextChat & 13 & 20{,}000 & 0.07\%  \\
Jupyter Notebook & 1{,}626 & 24{,}532 & 6.63\%  \\
ComfyUI & 357 & 6{,}241 & 5.72\%  \\
Gradio & 7 & 12{,}727 & 0.06\%  \\
vLLM & 3 & 948 & 0.32\%  \\
llama.cpp & 110 & 1{,}460 & 7.53\%  \\
GPT4All & 23 & 732 & 3.14\%  \\
Text Generation Web UI & 1 & 324 & 0.31\%  \\
Ray Serve & 2 & 188 & 1.06\%  \\
Llamafile & 42 & 412 & 10.19\%  \\
\bottomrule[1.2pt]
\end{tabular}
\begin{tablenotes}
    \footnotesize
    \item[1] With (w/) template response.
    \item[2] Without (w/o) template response.
\end{tablenotes}
\end{threeparttable}
}
\label{tab:response}
\end{table}
\begin{table*}[h!]
\centering
\caption{API Functionality Coverage Across LLM Deployment Frameworks.}
\label{tab:llm_api_functionality}
\resizebox{1\linewidth}{!}{
\begin{threeparttable}
\begin{tabular}{rcccccccccccc}
\toprule[1.2pt]
\textbf{Framework} & \textbf{Text/Chat Gen} & \textbf{Embed Gen} & \textbf{Img/Audio} & \textbf{Model Ops} & \textbf{File Ops} & \textbf{RAG} & \textbf{Fine-tune} & \textbf{Sess./Kernel} & \textbf{Task Queue} & \textbf{Sys Config} & \textbf{Moderation} & \textbf{App Deploy} \\
\midrule

\textbf{Ollama}             & \cellcolor{heatmap100}2 / 2\tnote{1} & \cellcolor{heatmap100}1 / 1 &  & \cellcolor{heatmap100}8 / 8 &  &  &  &  &  & \cellcolor{heatmap100}1 / 1 &  &  \\

\textbf{OpenWebUI}          & \cellcolor{heatmap0}0 / 2 & \cellcolor{heatmap0}0 / 1 &  & \cellcolor{heatmap66}1 / 2 & \cellcolor{heatmap0}0 / 1 & \cellcolor{heatmap0}0 / 3 &  &  &  &  &  &  \\

\textbf{Jan}                & \cellcolor{heatmap100}2 / 2 & \cellcolor{heatmap100}1 / 1 & \cellcolor{heatmap100}6 / 6 & \cellcolor{heatmap100}2 / 2 & \cellcolor{heatmap100}2 / 2 &  & \cellcolor{heatmap100}2 / 2 &  &  &  & \cellcolor{heatmap100}1 / 1 &  \\

\textbf{NextChat}           & \cellcolor{heatmap0}0 / 2 & \cellcolor{heatmap0}0 / 1 & \cellcolor{heatmap0}0 / 6 & \cellcolor{heatmap66}1 / 2 & \cellcolor{heatmap0}0 / 2 &  & \cellcolor{heatmap0}0 / 2 &  &  &  & \cellcolor{heatmap0}0 / 1 &  \\

\textbf{Jupyter Notebook}   &  &  &  &  & \cellcolor{heatmap0}0 / 6 &  &  & \cellcolor{heatmap0}0 / 13 &  & \cellcolor{heatmap100}1 / 1 &  &  \\

\textbf{ComfyUI}            &  & \cellcolor{heatmap100}1 / 1 & \cellcolor{heatmap66}1 / 2 & \cellcolor{heatmap100}2 / 2 & \cellcolor{heatmap66}1 / 2 &  & \cellcolor{heatmap66}1 / 2 & \cellcolor{heatmap100}1 / 1 & \cellcolor{heatmap100}1 / 1 &  & \cellcolor{heatmap100}1 / 1 \\

\textbf{Gradio}             & \cellcolor{heatmap0}0 / 1 &  &  &  & \cellcolor{heatmap0}0 / 1 &  &  &  & \cellcolor{heatmap0}0 / 2 & \cellcolor{heatmap66}1 / 2 &  &  \\

\textbf{vLLM}               & \cellcolor{heatmap66}1 / 2 & \cellcolor{heatmap100}1 / 1 & \cellcolor{heatmap0}0 / 6 & \cellcolor{heatmap0}0 / 2 & \cellcolor{heatmap100}2 / 2 &  & \cellcolor{heatmap66}1 / 2 &  &  &  & \cellcolor{heatmap0}0 / 1 &  \\

\textbf{llama.cpp}          & \cellcolor{heatmap90}4 / 5 & \cellcolor{heatmap100}1 / 1 &  & \cellcolor{heatmap66}1 / 2 &  &  &  &  &  & \cellcolor{heatmap100}1 / 1 &  &  \\

\textbf{GPT4All}            & \cellcolor{heatmap100}2 / 2 &  &  & \cellcolor{heatmap66}1 / 2 &  &  &  &  &  &  &  &  \\

\textbf{Text Gen WebUI}     & \cellcolor{heatmap66}2 / 3 & \cellcolor{heatmap100}1 / 1 & \cellcolor{heatmap100}6 / 6 & \cellcolor{heatmap66}3 / 4 & \cellcolor{heatmap100}2 / 2 &  & \cellcolor{heatmap33}1 / 2 &  &  &  & \cellcolor{heatmap100}1 / 1 &  \\

\textbf{Ray Serve}          &  &  &  &  &  &  &  &  &  &  &  & \cellcolor{heatmap33}1 / 3 \\

\textbf{Llamafile}          & \cellcolor{heatmap0}0 / 2 & \cellcolor{heatmap0}0 / 1 & \cellcolor{heatmap0}0 / 6 & \cellcolor{heatmap66}1 / 2 & \cellcolor{heatmap0}0 / 2 &  & \cellcolor{heatmap0}0 / 2 &  &  &  & \cellcolor{heatmap0}0 / 1 &  \\

\bottomrule[1.2pt]
\end{tabular}
\begin{tablenotes}
    \footnotesize
    \item[1] Each cell shows ``successful / total'' API endpoints in that category (e.g., \textit{2 / 2} means all responded successfully).
\end{tablenotes}
\end{threeparttable}
}
\end{table*}

The returned HTTP status codes also reflected security: some services returned 200 (full access), while many returned 401, 403, or 404, indicating authentication, access control, or endpoint obfuscation. Gradio-based deployments often returned 400, reflecting that the API was not designed for direct programmatic use. These response patterns reveal implicit security mechanisms across frameworks. High response rates are often associated with frameworks that expose APIs by default and lack built-in protection, while lower response rates generally correspond to more defensive configurations. It is worth noting that a significant portion of services respond with structured rejections (401/403), which suggests that some frameworks implement basic security measures even when deployed in public environments.

\subsubsection{Functional Coverage of Exposed Endpoints}

The functionality exposed through open API endpoints reveals not only the intended capabilities of LLM frameworks but also the extent to which internal operations are externally accessible, whether intentionally or not. As shown in \autoref{tab:llm_api_functionality}, the degree of exposure varies substantially across both frameworks and functionality categories.
Some frameworks expose broad functionality: for instance, Jan and llama.cpp each respond to over 10 distinct endpoints across text generation, embedding, file operations, and fine-tuning. ComfyUI returns valid responses for 11 categories, including session and queue management, reflecting its interactive, stateful design. In contrast, frameworks like OpenWebUI and Text Generation WebUI expose few usable endpoints despite implementing many, suggesting stricter access controls or incomplete external integration.
Across frameworks, the most commonly exposed functionality is text and chat generation, observed in 11 out of 13 frameworks. Model operations (e.g., listing and loading models) are also frequently reachable in 8 frameworks. However, sensitive features like fine-tuning, moderation, and knowledge base querying remain rare; for instance, only Jan and vLLM expose fine-tuning endpoints, and OpenWebUI supports RAG but with no responsive endpoints.
Exposure patterns highlight the trade-off between usability and security. Frameworks designed for local or development use often default to exposing internal APIs, lacking authentication or isolation, while production systems enforce stricter controls. Even non-critical endpoints, such as those for queue status or configuration, can inadvertently leak internal state.

\subsubsection{Per-Endpoint Result Representation}

All API responses are normalized into a unified schema encompassing  \textit{deployment framework}, \textit{endpoint category}, \textit{endpoint path}, \textit{response type}, and \textit{potential security relevance}, enabling consistent analysis across different frameworks. Based on this schema, we analyze Ollama as a representative case study. \autoref{tab:ollama} summarizes the endpoint responses for 155,424 observed calls.
\textbf{Text and chat completion} endpoints responded in only 0.19\% of cases, indicating limited exposure based on runtime state or configuration, but still potentially revealing system behavior. The embedding endpoint was accessible in 0.22\% of attempts; while rarely exposed, its exposure still poses a risk because embeddings can be used in inference attacks.
\textbf{Model management} endpoints showed uneven availability.Show model infomation and list running models were most frequently possible (9.94\%, 14.40\%), while listing local models was moderately possible (4.50\%), likely due to its role in coordination. Pull and push operations were low available (0.20\%, 0.19\%), while destructive actions like delete, copy, and create responded in fewer than 2\% of cases, suggesting partial lockdowns, though enforcement remains inconsistent.
\textbf{System-level} endpoints, including version and runtime model queries, responded in 14.48\% of cases, the same level to list running models. This likely reflects hardening efforts, but their occasional exposure highlights the need for stricter endpoint visibility controls.
The results highlight that in Ollama deployments, many endpoints are accessible without authentication, allowing unauthorized users to retrieve sensitive information and revealing a critical lack of access control.

\begin{table*}[t!]
\centering
\caption{Risk Categorization (Min $\sim$ Max Percentage) of Different LLM Deployment Frameworks.}
\label{tab:risk_categorization_sim}
\resizebox{1\linewidth}{!}{
\begin{threeparttable}
\begin{tabular}{rccccc}
\toprule[1.2pt]
\textbf{Framework} & \textbf{Model Info Disclosure} & \textbf{System Config Disclosure} & \textbf{Unauthorized \& Abuse} & \textbf{Vulnerabilities \& Reverse} & \textbf{Sensitive Content Gen} \\
\midrule

\textbf{ComfyUI}             & 5.26\% $\sim$ 10.54\%  & 5.64\%   & 5.06\% $\sim$ 5.26\%  & 5.35\% $\sim$ 10.09\%  & 5.06\%  \\

\textbf{Ollama}              & 0.00\% $\sim$ 14.48\% & --      & 0.19\% $\sim$ 1.28\% & --      & 0.19\% $\sim$ 0.22\% \\

\textbf{Text Gen WebUI}      & 0.29\%  & 0.29\%  & --      & --      & 0.29\%  \\

\textbf{GPT4All}             & 2.87\%  & --      & 5.10\%      & --      & --      \\

\textbf{Llamacpp}            & 3.36\% $\sim$ 6.99\%  & 6.71\%  & 11.92\%      & --      & --      \\

\textbf{Llamafile}           & 8.98\% $\sim$ 10.19\% & --      & 9.47\%      & --      & --      \\

\textbf{Jupyter Notebook}    & --      & 8.06\%      & --      & 0.16\% & --      \\

\bottomrule[1.2pt]
\end{tabular}

\begin{tablenotes}
    \footnotesize
    \item[1] Each cell shows the observed minimum $\sim$ maximum percentage of API endpoints posing the corresponding risk category. If minimum equals maximum, only a single value is shown. ``--'' indicates no relevant risk observed. Detailed case rates for ComfyUI endpoints are summarized in \autoref{tab:comfyui_risk_table}.
\end{tablenotes}
\end{threeparttable}
}
\end{table*}
\begin{takeawaybox}
\textbf{Answer to RQ2:} \textit{API exposure and access control vary across LLM frameworks. Ollama and Llamafile responded to over 10\% of unauthenticated requests, while other frameworks, such as Open WebUI and Gradio, responded at rates below 0.1\%. Most frameworks exposed basic text generation functionality, but some also leaked model or system information, particularly Ollama, where this rate exceeded 14\%. Many frameworks relied on permissive defaults, exposing deployments to the risk of unauthorized access.}
\end{takeawaybox}

\begin{table}[t!]
\centering
\caption{Responsiveness of Ollama API Endpoints.}
\resizebox{1\linewidth}{!}{
\begin{threeparttable}
\begin{tabular}{cllrr}
\toprule[1.2pt]
\textbf{Category} & \textbf{Endpoint} & \textbf{Function} & \textbf{\# Resp.} & \textbf{Resp. Rate}\tnote{1} \\
\midrule
\multirow{2}{*}{Text/Chat Gen}
    & /api/generate & Text Completion         & 299 & 0.19\% \\
    & /api/chat     & Chat Completion         & 300 & 0.19\% \\
\midrule
\multirow{1}{*}{Embedding}
    & /api/embeddings & Generate Embedding     & 343  & 0.22\% \\
\midrule
\multirow{8}{*}{Model Ops}
    & /api/tags     &   List Local Models            & 7{,}001 & 4.50\% \\
    & /api/pull     &   Pull Model          & 312 & 0.20\% \\
    & /api/push     &  Push Model          & 299 & 0.19\% \\
    & /api/delete   &  Delete Model            & 306 & 0.20\% \\
    & /api/copy     &  Copy Model        & 1986  & 1.28\% \\
    & /api/show     &  Show Model Info           & 15{,}452  & 9.94\% \\
    & /api/create   &  Create Model     & 311  & 0.20\% \\
    & /api/ps       &  List Running Models     & 22{,}375  & 14.40\% \\
    & /api/running  & List Loaded Models      & 0   & 0.00\% \\
\midrule
\multirow{1}{*}{System Config}
    & /api/version  & Get Ollama Version      & 22{,}503   & 14.48\% \\
\bottomrule[1.2pt]
\end{tabular}
\begin{tablenotes}
    \footnotesize
    \item Note: Based on 155{,}424 sampled LLM service invocations.
\end{tablenotes}
\end{threeparttable}
}
\label{tab:ollama}
\end{table}

\subsection{RQ3: Security and Risk Analysis}

As shown in \autoref{tab:risk_categorization_sim}, this section systematically analyzes potential security risks in LLM deployment frameworks. 
Each subcategory combines observed cases with deeper security insights to reveal not only surface-level exposures but also underlying attack vectors and systemic vulnerabilities. 

\subsubsection{Model Information Disclosure}

Model information disclosure is among the most prevalent security risks across LLM deployment frameworks, as indicated in \autoref{tab:risk_categorization_sim}. For example, the \texttt{/show} endpoint can reveal detailed model metadata, while the \texttt{/history} endpoint may leak prior inference workflows, outputs, and fine-tuning traces, compromising user privacy and exposing proprietary data. Notably, in Ollama, the \texttt{/api/show} endpoint exhibits an exceptionally high exposure rate of 14.40\%, highlighting the severity of this risk even in widely adopted deployment platforms.
The \texttt{/embeddings} endpoint (5.26\% exposure) in ComfyUI further illustrates this problem by exposing the list of loaded textual inversion embeddings, such as \textit{Bad-Hands-XL}, inadvertently revealing deployment purposes. Similarly, in \texttt{llama.cpp}, the \texttt{/v1/models} endpoint may disclose detailed model information, including the actual deployment paths of \texttt{.gguf} model files, significantly amplifying the risk of targeted attacks.
Such leakage not only facilitates customized attacks but also heightens the risks of model exploitation, prompt injection, and unauthorized access. In severe cases, attackers could reconstruct user intents, manipulate outputs, or compromise the deployment’s integrity and confidentiality.

\begin{table}[t!]
\centering
\caption{Security Risk Exposure of ComfyUI Endpoints.}
\label{tab:comfyui_risk_table}
\resizebox{1\linewidth}{!}{
\begin{threeparttable}
\begin{tabular}{clrr}
\toprule[1.2pt]
\textbf{Security Risk} & \textbf{API Endpoint} & \textbf{\# Cases} & \textbf{Case Rate\tnote{1}} \\
\midrule

\multirow{2}{*}{Model Information Disclosure}
    & \texttt{/embeddings} & 330  & 5.29\% \\
    & \texttt{/history}    & 658 & 10.54\% \\

\midrule

\multirow{1}{*}{System Configuration Disclosure}
    & \texttt{/system\_stats} & 352 & 5.64\% \\

\midrule

\multirow{2}{*}{Unauthorized \& Resource Abuse}
    & \texttt{/queue}  & 316  & 5.26\% \\
    & \texttt{/prompt} & 328 & 5.06\% \\

\midrule

\multirow{2}{*}{Vulnerabilities \& Reverse}
    & \texttt{/object\_info} & 630  & 10.09\% \\
    & \texttt{/extensions}   & 334 & 5.35\% \\  

\midrule

\multirow{1}{*}{Sensitive Content Generation}
    & \texttt{/prompt} & 328 & 5.06\% \\

\bottomrule[1.2pt]
\end{tabular}
\begin{tablenotes}
    \footnotesize
    \item [1] Based on analysis of 6{,}241 sampled ComfyUI endpoints. \# Cases and Case Rates indicate confirmed security risks.
\end{tablenotes}
\end{threeparttable}
}
\end{table}

\subsubsection{Hardware and System Configuration Disclosure}

Disclosure of system configuration details further enlarges the attack surface by providing adversaries with valuable intelligence about the underlying environment. ComfyUI’s \texttt{/system\_stats} endpoint (5.64\%) reveals operating system types, total and available memory, GPU specifications (e.g., RTX 4090 with 24GB VRAM), and ComfyUI version information.
Such granular system insights enable attackers to craft hardware-specific resource exhaustion attacks, such as GPU memory flooding, or identify platform-specific vulnerabilities (e.g., Windows NT kernel exploits). Alarmingly, among instances exposing \texttt{/system\_stats}, 41.28\% were found to be publicly accessible via \texttt{--listen 0.0.0.0} without any authentication mechanisms. When combined with the presence of high-performance GPUs, such exposures create ideal conditions for unauthorized resource exploitation, including GPU hijacking, cryptocurrency mining, large-scale model inference, or persistent backdoor implantation. 

\subsubsection{Unauthorized Access and Resource Abuse}

Unauthorized access and resource abuse represent direct threats to the availability and stability of deployed services. Endpoints such as \texttt{/queue} (5.26\%) and \texttt{/prompt} (5.06\%) allow unauthenticated users to submit inference tasks or monitor system load.
The ability to observe task queues enables attackers to perform real-time load sensing, identifying idle periods when resource-draining or prompt injection attacks can be launched with minimal resistance. Furthermore, submitting crafted prompt graphs to \texttt{/prompt} without authentication exacerbates the risk: attackers could overload the system with computationally expensive tasks, rapidly depleting GPU memory and causing service outages (denial-of-service attacks). By continuously monitoring queue states, attackers can also infer operational patterns over time, improving the precision and persistence of their abuse strategies. In the absence of fine-grained access control and rate-limiting, such systems remain highly vulnerable to long-term degradation and operational hijacking.

\subsubsection{Vulnerabilities and Reverse Engineering}

The exposure of internal modules and metadata creates direct opportunities for vulnerability discovery and reverse engineering. In platforms like ComfyUI, endpoints such as \texttt{/extensions} (5.35\%) and \texttt{/object\_info} (10.09\%) reveal detailed information about installed nodes, system structure, and behaviors. With such insights, attackers can reconstruct internal workflows, pinpoint critical components like \texttt{PythonEvalNode}, and exploit weaknesses such as insecure APIs or insufficient validation. Nodes handling external communication (e.g., \texttt{GeminiAPINode}) may leak API keys, while backend modules (e.g., \texttt{cm-api.js}) expose surfaces for command injection.
Similar risks arise in another framework. Jupyter Notebook instances exposing the \texttt{/api} endpoint disclose version data (e.g., ``5.5.0''), allowing attackers to link targets to known vulnerabilities, including unauthorized API access, unauthenticated WebSocket RCE, and token bypass. Fingerprinting deployments and mapping them to public exploits significantly accelerates attack development.
Passive metadata collection can quickly escalate into full system compromise, underscoring the critical risks of seemingly minor exposures.

\subsubsection{Sensitive Content Generation}

Sensitive content generation presents a significant risk, particularly in deployments lacking robust input validation and output moderation. Platforms such as Ollama and Text Generation WebUI expose endpoints that accept custom prompts or generation parameters, which attackers can exploit to induce the production of offensive, illegal, or otherwise sensitive outputs. In Ollama, LLMs often labeled as ``uncensored'', can be manipulated through crafted prompts to bypass moderation controls. Similarly, in Text Generation WebUI, insufficient prompt sanitization may allow adversarial inputs to elicit harmful responses or extract proprietary system behaviors.
If prompt histories, session contexts, or interaction logs are improperly secured, attackers can retrieve past prompts and outputs to refine injection strategies and escalate misuse. Such manipulations may cause reputational harm, legal exposure, and regulatory penalties. As scrutiny of AI-generated content grows, securing the content generation pipeline is essential to mitigating systemic risks.

\begin{takeawaybox}
\textbf{Answer to RQ3:} \textit{Security risk analysis shows that model information disclosure, system leaks, unauthorized access, vulnerabilities, and sensitive content generation are prevalent across LLM deployment frameworks, though unevenly. Notably, frameworks like ComfyUI expose endpoints across all major risk categories, indicating broad and systemic weaknesses. The persistence of such exposures indicates that insecure deployment practices are widespread, and securing LLM systems demands rethinking default configurations, strengthening access controls, and minimizing exposure surfaces beyond patching individual flaws.}
\end{takeawaybox}

\section{Discussion}
\label{sec:discussion}

\subsection{Recommendations and Best Practices}

Improving the security posture of local LLM frameworks requires coordinated efforts across tooling, usage, and community practices.
\textbf{For tool developers.}
Frameworks should adopt a secure-by-default approach. APIs should bind to \texttt{localhost} by default, require explicit configuration for external access, and issue warnings when sensitive endpoints (e.g., model deletion, file upload) are accessed without authentication. Interface metadata, such as OpenAPI schemas, can define access roles and scopes, enabling future integration with RBAC systems. Optional security features should be documented clearly and made easier to configure.
\textbf{For end users.}
Users should avoid assuming that ``local'' tools are secure by default. Exposing endpoints, including those within containers or LAN environments, should be done with caution. Recommended practices include restricting interfaces, enabling authentication where available, and disabling unnecessary endpoints. Users should consult community guides and treat LLM tools as potentially network-accessible services.
\textbf{For the broader ecosystem.}
The blurred line between research prototypes and production deployments poses systemic risks. Insecure defaults in upstream tools can propagate into downstream projects, widening the attack surface. Community resources such as deployment checklists, secure-by-default templates, and best practices can help align tool design with real-world deployment needs. Security should be integral to LLM system design, not treated as an afterthought.

\subsection{Threats to Validity}
\noindent\textbf{Deployment Framework Coverage.} 
Our study focuses on a limited set of LLM deployment frameworks. We analyzed major platforms such as ComfyUI, Ollama, and Text Generation Web UI. Nevertheless, the insecure practices identified, including sensitive data exposure, unauthorized access, and resource abuse, reflect systemic issues common to LLM deployments. These risks are not limited to the studied frameworks and are broadly applicable across the ecosystem.

\noindent\textbf{Sampling Coverage.}
Given the identification of over 320,000 exposed services, exhaustive probing would have imposed undue load on platform servers. To ensure ethical sampling while maintaining statistical rigor, we adopted a 95\% confidence level with a 5\% margin of error. Although representative, our sample may not capture certain rare or customized environments, which future work could further explore.

\subsection{Ethical Considerations}

All activities are carefully designed to minimize impact and respect the privacy of service operators. We probe only publicly accessible endpoints using non-destructive, read-only API requests, without attempting to bypass authentication, access private data, or exploit vulnerabilities. Sensitive findings are handled responsibly following coordinated vulnerability disclosure practices when appropriate. The study follows institutional and legal ethical guidelines for Internet research and collects no personally identifiable information.

\section{Conclusion}
\label{sec:conclusion}
This work presents a comprehensive view of the current landscape of public-facing LLM deployments, offering the first large-scale empirical evidence across 320,102 services spanning 15 frameworks. Our analysis highlights both the rapid expansion and decentralization of self-hosted LLM ecosystems and the widespread presence of security risks, including model disclosure, system configuration leakage, unauthorized access and resource abuse, vulnerabilities, and sensitive content generation. These findings reveal critical gaps between deployment practices and security requirements. Moving forward, we will extend our study with temporal analysis of deployment trends and more detailed assessments of security and operational risks in real-world LLM ecosystems.

\bibliographystyle{IEEEtranS}
\bibliography{main}

\end{document}